\documentclass[11pt,a4]{article}
\usepackage[pdftex]{hyperref}
\usepackage{amsthm,amsmath,latexsym,amssymb,amsfonts,color,calrsfs}
\usepackage{graphicx,lscape,fancyhdr,array,stmaryrd,euscript,wrapfig}
\newcommand{\be}{\begin{equation}}
\newcommand{\ee}{\end{equation}}
\newcommand{\bee}{\begin{eqnarray}}
\newcommand{\beee}{\begin{array}}
\newcommand{\eee}{\end{eqnarray}}
\newcommand{\eeee}{\end{array}}
\definecolor{rougef}{rgb}{0.56,0,0}
\definecolor{vertf}{rgb}{0,0.5,0}
\definecolor{bleuf}{rgb}{0,0,0.8}

\begin{document}
\begin{titlepage}

\begin{flushright}
\vspace{1mm}
\end{flushright}

\vspace{1cm}
\begin{center}
{\bf \Large  Frame-like off-shell dualisation for mixed-symmetry gauge fields}
\vspace{2cm}

\textsc{Nicolas Boulanger\footnote{Research Associate of the Fund
for Scientific Research-FNRS (Belgium);
nicolas.boulanger@umons.ac.be} and Dmitry
Ponomarev\footnote{dmitri.ponomarev@umons.ac.be}}

\vspace{2cm}

{\em Service de M\'ecanique et Gravitation, Universit\'e de Mons -- UMONS\\
20 Place du Parc, 7000 Mons (Belgium)}

\end{center}

\vspace{0.5cm}
\begin{abstract}

We construct a purely frame-like parent action that allows to dualise, 
at the off-shell level, an arbitrary 
mixed-symmetry bosonic massless fields in Minkowski background of dimensions $d\,$. 
Starting from any massless mixed-symmetry gauge field in the standard Skvortsov 
frame-like formulation and following an off-shell dualisation procedure, 
we obtain dual theories which are on-shell related by $so(d-2)$ Hodge duality. 
The Hodge dualisation can be done on any column of the 
Young diagram characterizing the generalised spin
of the original frame-like field.
Dualisation with respect to the first column of the Young diagram 
leads to a standard frame-like action for the dual field.
Any other dualisation results in an action 
which cannot be described by the standard frame-like formalism, 
as the on-shell field is not $so(d-2)$ traceless.
Instead, the latter field is given by the product of an irreducible traceless
tensor and a certain number of $so(d-2)$-invariant metrics,  
and the corresponding dual frame-like action is new. 
Such actions require supplementary fields, 
which naturally arise along the lines of the approach that we propose.

\end{abstract}

\end{titlepage}

\tableofcontents

\numberwithin{equation}{section}

\section{Introduction}
\label{intro}

Mixed-symmetric gauge fields have attracted a lot of attention these recent years, 
partly because the totally-symmetry case is fairly well understood by now 
even at the full nonlinear level \cite{Vasiliev:1990cm,Vasiliev:1992av,Vasiliev:2003ev}, 
but also because in dimension higher than 4, mixed-symmetry fields are allowed from the 
point of view of representation theory of the corresponding spacetime 
isometry algebra. They also appear in string theory, albeit at the
massive level, see e.g. \cite{Francia:2006hp} for related discussions.
\vspace*{.3cm}

In the context of string-field theory, mixed-symmetry fields were studied
in the eighties and Lagrangians in flat space were explicitly given for
some cases in a metric-like fashion, see e.g. \cite{Curtright:1980yk,Aulakh:1986cb}. 
Using string-field-like techniques,  
Labastida \cite{Labastida:1986gy,Labastida:1987kw} proposed 
a Lagrangian describing an arbitrary free, ${gl}(d)$-irreducible, 
mixed-symmetry gauge field in flat background.
It was proved much later \cite{Bekaert:2003az,Bekaert:2006ix} 
that the corresponding theory indeed propagates the correct degrees of freedom.
Still in a flat background and for metric-like fields, more recent works can
be found in  \cite{Campoleoni:2008jq,Campoleoni:2009gs}
where, among various results, the equivalent of the Labastida action
but for arbitrary tensor-spinor fields was obtained.
\vspace*{.3cm}

An achievement was done within the frame-like and unfolded 
approach when Skvortsov took advantage of the Cartan formulation of gauge theories in 
order to describe, both on-shell \cite{Skvortsov:2008vs} 
and off-shell \cite{Skvortsov:2008sh}, 
arbitrary mixed-symmetry gauge fields freely propagating in flat 
spacetime. Fermionic field are treated along the same lines in \cite{Skvortsov:2010nh}. 
See \cite{Bekaert:2002dt,deMedeiros:2002ge,Zinoviev:2002ye,Zinoviev:2003ix,Buchbinder:2007ix,Moshin:2007jt,Alkalaev:2008gi,Fotopoulos:2008ka,Zinoviev:2009vy,Buchbinder:2011xw} 
and references therein for more works on mixed-symmetry gauge fields in flat background. 
\vspace*{.3cm}

Mixed-symmetry gauge fields can also appear via dual formulations of totally symmetric fields
\cite{Curtright:1980yj,Casini:2001gv,West:2001as,
Hull:2001iu,Bekaert:2002dt,deMedeiros:2002ge,Casini:2003kf,Bekaert:2003az,
Boulanger:2003vs,Ajith:2004ia,Zinoviev:2005zj,Zinoviev:2005qp}.
An off-shell and covariant description of the double-dual graviton,  
a field first introduced in \cite{Hull:2001iu}, 
was obtained in a recent paper \cite{Boulanger:2012df}.  
This was done in the metric-like approach, and 
the purpose of the present paper is to give a frame-like treatment of the off-shell 
dualisation procedure, thereby allowing us to treat the arbitrary mixed-symmetric 
case via the frame-like formulation \cite{Skvortsov:2008vs,Skvortsov:2008sh}.
\vspace*{.3cm}

The plan of the paper is as follows. 
In the next section \ref{umsf} we briefly review some basics of 
unfolding and related issues. 
In section \ref{sec:metric} we review the off-shell dualisation 
of linearised gravity on the first column of the gauge field, 
in the metric-like formalism.
Then, in section \ref{fldg} we translate the previous analysis to 
the frame-like approach. We then perform the second nontrivial
off-shell dualisation of the graviton in a frame-like and first-order fashion. 
The latter analysis is generalised to the arbitrary mixed-symmetry case
in section \ref{sec:mixed}.
Finally, we give some conclusions and perspectives in section \ref{sec:concl}, 
followed by an appendix summarizing our notation. 

\section{Unfolding mixed-symmetry fields}
\label{umsf}

In this Section we briefly review some basic concepts concerning 
the unfolded approach \cite{Vasiliev:1988xc,Vasiliev:1988sa} 
in general  as well as the unfolded formulation for massless 
mixed-symmetry fields in flat background developed in \cite{Skvortsov:2008vs,Skvortsov:2008sh}.
\vspace*{.3cm}

Unfolding means reformulation of the theory in terms of 
differential form fields $\{W^{\alpha}(x)\}_{\alpha\in{\cal S}}$
\footnote{For a dynamical system propagating local degrees of freedom, 
the set ${\cal S}$ is infinite-dimensional 
due to the presence of infinitely-many zero-form fields.}
subjected to generalized zero curvature conditions 
\begin{equation}
\label{unf1}
R^{\alpha}:=dW^{\alpha}+G^{\alpha}(W(x))=0\;,
\end{equation}
where $d$ denotes the exterior derivative and $G^{\alpha}(W)$ are 
wedge-product polynomials in the $W^{\alpha}$'s, with 
the wedge product used implicitly throughout this paper, 
\begin{equation}
\label{unf2}
G^{\alpha}(W):=\sum_{n=1}^{\infty}
f^{\alpha}_{\beta_1\dots\beta_n}W^{\beta_1} \dots W^{\beta_n}\;,
\end{equation} 
and the indices $\alpha$ are some collective fiber indices, 
e.g. fiber Lorentz indices. 
In addition $G^{\alpha}(W)$ satisfies the integrability 
condition
\begin{equation}
\label{unf3}
G^{\alpha}(W)\frac{{{\partial}} G^{\beta}(W)}{{{\partial}} W^{\alpha}} \equiv 0\;,
\end{equation}
which just states that (\ref{unf1}) is compatible with $d^2=0\,$.
All the dynamical information about unfolded system is encoded 
in the set of fields $W^{\alpha}$ used as well as in the 
structure constants $f^{\alpha}_{\beta_1\dots\beta_n}$
(\ref{unf2}).
\vspace*{.3cm}

In the same manner as each field $W^{\alpha}_{\bf p_{\alpha}}$ 
of differential form degree $p_{\alpha}$ has an associated
curvature $R_{\bf p_{\alpha}+1}^{\alpha}$
of differential form degree $p_{\alpha}+1$ 
(\ref{unf1}),
each field with $p_{\alpha}\geqslant 1$ has an associated gauge 
parameter $\varepsilon_{\bf p_{\alpha}-1}^{\alpha}$ of
differential form degree $p_{\alpha}-1\,$. 
The invariance of  (\ref{unf1}) with respect to the gauge transformations 
\begin{equation*}
\delta W^{\alpha}=d\varepsilon^{\alpha}-
\varepsilon^{\beta}\frac{{{\partial}} G^{\alpha}(W)}{{{\partial}} W^{\beta}} \quad 
\text {for} \quad p_{\alpha}\geqslant 1\;,
\end{equation*}
\begin{equation}
\label{unf4}
\delta W^{\alpha}=-
\varepsilon^{\beta}\frac{{{\partial}} G^{\alpha}(W)}{{{\partial}} W^{\beta}} \quad 
\text {for} \quad p_{\alpha}=0\;,
\end{equation}
is manifest due to the condition (\ref{unf3}).
The same condition guarantees the generalized Bianchi identity
\begin{equation}
\label{unf6}
dR^{\alpha}-R^{\beta}\frac{{{\partial}} G^{\alpha}}{{{\partial}} W^{\beta}}\equiv 0\;.
\end{equation}
\vspace*{.1cm}
 
One can apply the same principle to the 
unfolded-like equations $\delta W^{\alpha}=0$ 
on the gauge parameters $\varepsilon^{\alpha}\,$. 
They also possess manifest gauge symmetries 
generated by parameters $\bar\varepsilon_{\bf p_{\alpha}-2}^{\alpha} $, 
each of them associated with some gauge parameter 
$\varepsilon_{\bf p_{\alpha}-1}^{\alpha}$
with $p_{\alpha}\geqslant 2\,$. 
The gauge parameters $\bar\varepsilon^{\alpha}$
give rise to the second order gauge transformations 
associated with (\ref{unf1}).
\vspace*{.3cm}
  
Continuing this line of reasoning one can find that each field 
$W_{\bf p_{\alpha}}^{\alpha}$
is accompanied with a chain of $p_{\alpha}$ different-level gauge parameters 
of decreasing differential form degrees from $p_{\alpha}-1$ down to zero:
\begin{equation}
\label{unf5}
W_{\bf p_{\alpha}}^{\alpha}\quad \rightarrow 
\quad \varepsilon_{\bf p_{\alpha}-1}^{\alpha} \quad
\rightarrow \quad \bar\varepsilon_{\bf p_{\alpha}-2}^{\alpha} 
\quad \rightarrow \quad \dots\,.
\end{equation} 
Hence, a formulation of the theory in the unfolded form makes manifest  
all the gauge symmetries together with all their reducibilities. 
The requirement that all the symmetries be manifest 
uniquely determines the unfolded equations for massless mixed-symmetry fields. 
\vspace*{.3cm}

To unfold any theory in Minkowski background one should first
describe Minkowski space in the unfolded form and then add matter fields 
subjected to their equations of motion such that the whole system 
remains compatible. 
The background Minkowski space can be described via the zero curvature 
condition for a one form $\Omega_0=h^aP_a+\varpi^{ab}M_{ab}$ valued 
in the Poincar\'e algebra generated by the translations $P_a$ and 
the Lorentz algebra generators $M_{ab}$
\begin{eqnarray}
&{\cal R}:=d\Omega_0+\Omega_0\Omega_0 = 0\;, \quad
{\cal R}\equiv T^aP_a+R^{ab}M_{ab} \quad &
\nonumber \\
\label{background}
&\Rightarrow  \quad T^a=dh^a+\varpi^{a,}{}_bh^b=0\;, \quad R^{ab}=
d\varpi^{ab}+\varpi^{a,}{}_c\varpi^{cb}=0\;,&
\end{eqnarray}
where $h^a$, $\varpi^{ab}=-\varpi^{ba}$, $T^a$ and $R^{ab}=-R^{ba}$ 
are the vielbein, the spin-connection, the torsion and the Riemann curvature,  
respectively. We assume that $h_\mu^a$ is a non-degenerate matrix, so it can 
be used to transform base indices to fiber ones and vice versa. 
Eqs. (\ref{background}) are unfolded because the associated 
integrability condition (\ref{unf3}) is fulfilled as a consequence 
of the Jacoby identity for the Poincar\'e algebra. We will use 
Cartesian coordinates $h^a_{\mu}=\delta^a_{\mu}$, $\varpi^{ab}=0$ in 
what follows. This allows to treat base and fiber indices on the same footing.
\vspace*{.3cm}

As for any non-trivial unitary representation of the non-compact Poincar\'e
group, the representations carried by massless mixed-symmetry fields 
are infinite-dimensional.  
In general, the unfolded formulation for such  systems  
requires infinite number of fields subjected to infinite 
number of equations of motion.
To make contact with ordinary field-theoretical approaches 
one should find which fields and equations are dynamical.
The remaining non-dynamical fields are either auxiliary, 
\emph{i.e.} expressible as derivatives of the dynamical fields, 
or Stueckelberg-like, meaning that they can be gauged away by 
algebraic gauge symmetries. 
Analogously, non-dynamical equations are either constraints, 
\emph{i.e.} equations that are satisfied identically when 
the auxiliary fields are expressed in terms of the dynamical ones, 
or consequences of dynamical equations, thereby not 
imposing any further restriction on the dynamical fields.
\vspace*{.3cm}

It will be convenient to formally extend  the Young diagram 
$\mathbf{Y}[h_1,h_2, \dots, h_{s_1}]$ by
the infinite number of columns of zero heights 
$h_i=0$ for $i>s_1$. This gives rise to 
an infinite sequence of non-increasing non-negative 
integer numbers $h_i$.
It was shown in \cite{Skvortsov:2008vs} that for 
a massless spin-$\mathbf{Y}$ field freely propagating 
in the Minkowski space, the unfolded formulation requires a set of fields
$W_{\mathbf{p^g}}^{\mathbf Y^g}$ enumerated by a positive integer $g\,$,
the fields $W_{\mathbf{p^g}}^{\mathbf Y^g}$ being differential $p^g$-form 
taking their values in traceless $\mathbf Y^g$-shaped $so(d-1,1)$ 
tensors\footnote{Here and in the following, 
when we discuss irreducible representations of $so(m,n)\,$,  
we actually do not consider (anti) self-duality conditions.}. 
The differential form degrees $p^g$ and the shapes $\mathbf Y^g[h^g_1,h_2^g,\dots]$ 
are defined by the generalized spin $\mathbf{Y}[h_1,h_2,\dots]$ in the following way
\begin{equation}
\label{unfspace}
p^g=h_g\;, \qquad 
\begin{array}{rclcl}
h_i^g &= & h_i+1 & \text{for} & i<g\\
 {} & {} & {} & {} & {}\\
h_i^g &=& h_{i+1} & \text{for} &  i \geqslant g\;.\\
\end{array}
\end{equation}
Since both the differential form degree and the fiber space representations are 
uniquely determined by the fields grade $g$, we will often write $W^g$ instead 
of $W_{\mathbf{p^g}}^{\mathbf Y^g}\,$. Similarly, the associated curvatures and 
gauge parameters will be denoted by $R^g$ and $\varepsilon^g\,$, respectively.
\vspace*{.3cm}

The unfolded equations take the form
\begin{equation}
\label{unfmix}
R^g := dW^g + \sigma_-(h) W^{g+1} = 0\;,
\end{equation}
where $\sigma_-(h)$ is an operator built out of $p^{g}-p^{g+1}+1$ 
background vielbeins $h^a$ and mapping fiber-space
$\mathbf Y^{g+1}$-shaped traceless tensors
to $\mathbf Y^{g}$-shaped traceless tensors,
which defines it up to unessential overall factor.
The $\sigma_-$ operator carries the index ``$-$" due to the 
fact that it decreases the field grade $g$. 
The integrability condition implies $\sigma_-^2=0\,$.
\vspace*{.3cm}

The manifest gauge symmetries (\ref{unf4}) for (\ref{unfmix}) 
acquire the form 
\begin{equation}
\label{unfmix1}
\delta W^g=
d\varepsilon^g+
(-1)^{p^{g}-p^{g+1}}
\sigma_-(h)\varepsilon^{g+1}\;.
\end{equation}
Each $W^g$ has an associated chain 
of $p^g$ gauge parameters of different levels of reducibility. 
The gauge transformations of higher reducibility levels are of the same form 
as (\ref{unfmix}).
The Bianchi identities are
\begin{equation}
\label{unfmix2}
dR^g+(-1)^{p^{g}-p^{g+1}}
\sigma_-(h)R^{g+1}\equiv 0\;.
\end{equation}

The analysis of the  unfolded equations  amounts
to ${\rm H}(\sigma_-)$ computations \cite{Lopatin:1987hz, Shaynkman:2000ts}, 
see also \cite{Ponomarev:2010ji} for recent developements,  
and goes as follows. 
The fields can be divided into three groups:
\begin{itemize}
\item the $\sigma_-$-exact fields that  
 can be gauged away by the Stueckelberg gauge symmetries, 
 the second term on the right-hand side of (\ref{unfmix1});
 \item the fields that are not $\sigma_-$-closed and  
 can therefore be expressed in terms of the lower-grade fields via (\ref{unfmix}). 
 They are auxiliary;
 \item the remaining fields that belong to ${\rm H}(\sigma_-)$. They are the dynamical fields. 
 \end{itemize} 
Similarly, one can split the curvatures and the associated equations 
$R^g=0$ into the following groups:
\begin{itemize}
\item the projection of the equation (\ref{unfmix}) to its
 $\sigma_-$-exact component expresses the grade-$(g+1)$ 
 field $\sigma_-W^{g+1}$ 
in terms of the first derivatives of the grade-$g$ field $W^g\,$.
The $\sigma_-$-exact component of $R^g=0$ therefore is a constraint;
\item from (\ref{unfmix2}) it follows that once the equation 
$R^g=0$ has been taken into account,  
it enforces the part of $R^{g+1}$ that is not annihilated by $\sigma_-$ 
to be zero as a consequence;
\item the remaining curvatures belong to ${\rm H}(\sigma_-)$ and 
give rise to the dynamical equations.
\end{itemize} 
Let us note that the fields and the curvatures are valued in the same fiber 
spaces but carry different differential form degrees: $p^g$ for 
grade-$g$ fields and $p^{g}+1$ for grade-$g$ curvatures. 
So, looking for dynamical fields and dynamical equations one 
should compute $\sigma_-$-cohomologies in different differential 
form degrees. It can be shown analogously that the ${\rm H}(\sigma_-)$ 
for differential form degrees less then $p_g$ define 
differential gauge symmetries of different levels, while the 
${\rm H}(\sigma_-)$ for differential form degrees higher 
then $p_g+1$ define associated Bianchi identities. See also 
\cite{Boulanger:2008up} for related comments and extended discussions 
concerning the zero-form sector. A master-field reformulation 
of Skvortsov's equations can be found in \cite{Boulanger:2008kw}. 
\vspace*{.3cm} 
   
Rigorous computations show \cite{Skvortsov:2008vs} that for the unfolded equations 
(\ref{unfspace}), (\ref{unfmix}) the only dynamical field $\varphi$ belongs 
to $W^1$. It can be identified with the Labastida 
metric-like field. The first equation 
\begin{equation}
\label{an1}
R^1:=dW^1+\sigma_-W^2=0
\end{equation}
does not impose equations 
on dynamical field, just expressing the first auxiliary field $W^2$ 
in terms of the first derivative of the dynamical one. 
The only dynamical equation is a certain trace projection 
of the second equation
\begin{equation}
\label{an2}
\text{Tr}(R^2):=\text{Tr}(dW^2)=0
\end{equation} 
(the trace projection is such that it annihilates $\sigma_-W^3$).
Substituting $W^2$ expressed in terms of $\varphi$ into 
(\ref{an2}) one gets the second order differential equation for $\varphi\,$.
\vspace*{.3cm} 
 
One can think of $W^1$ and $W^2$ as generalizations of 
the vielbein and the spin-connection of the Cartan formulation of gravity to 
the mixed-symmetry case. Then (\ref{an1}) generalizes the 
zero-torsion constraint, while (\ref{an2}) generalizes 
the Einstein equations. Exploiting this analogy we denote 
$W^1$, $W^2$, $R^1$ and $R^2$ by $e$, $\omega$, $T$ and $R$ respectively.
Let us also denote $\varepsilon^1$ by $\xi$ and $\varepsilon^2$ by 
$\lambda$.
\vspace*{.3cm} 
 
The shapes $\mathbf{Y}^1[h_2,h_3,\dots]$ and $\mathbf{Y}^2[h_1+1,h_3, \dots]$
differ only by the heights of the first column. 
For a degree $p$-form $\Phi_{\mathbf{p}}^{\mathbf{Y}^1}$ and a degree $q$-form 
$\Psi_{\mathbf{q}}^{\mathbf{Y}^2}$ provided that $p+q=h_1+h_2+1\,$,  
there is a unique scalar product 
\begin{equation}
\label{scprod}
\langle\Phi_{\mathbf{p}}^{\mathbf{Y}^1}|\Psi_{\mathbf{q}}^{\mathbf{Y}^2} \rangle=
\int{\Phi_{\mathbf{p}}^{a[h_2],b[h_3],\dots}
\Psi_{\mathbf{q}}^{a[h_1+1]}{}_{,b[h_3],\dots}}H_{a[h_1+h_2+1]}\;.
\end{equation}
As shown in \cite{Skvortsov:2008sh} the first order action 
\begin{equation}
\label{action}
S=\langle de+\frac{1}{2}\sigma_-\omega|\omega\rangle
\end{equation}
is a unique action, which is invariant under (\ref{unf4}) and 
free of derivatives higher than two.
The variation of the action gives
\begin{equation}
\label{action1}
\frac{\delta S}{\delta \omega}=\pi_2[T]=0,
\end{equation}
\begin{equation}
\label{action2}
\frac{\delta S}{\delta e}=\pi_1[R]=0,
\end{equation}
where $\pi_1$ and $\pi_2$ are the projectors induced by the contractions 
of $R$ and $T$ with $e$ and $\omega$ respectively. 
Since $R$ has two indices more than $e$, $\pi_1$ takes one trace.
Let us note that in general $\pi_2$ is not invertible, hence 
(\ref{an1}) does not follow from (\ref{action1}).
However, (\ref{action1}) proves to be 
sufficient in order to express the components of $\omega$ 
contributing to the dynamical equation 
(\ref{action2}) in terms of the dynamical field $\varphi\,$.
By exploiting the Stueckelberg gauge symmetries, expressing 
the auxiliary field $\omega$ in terms of the dynamical field 
$\varphi$ and plugging the result back into the action (\ref{action}), 
one recovers the Labastida metric-like second-order action.
\vspace*{.3cm} 

It is easy to see that all the fields of the unfolded system (\ref{unfspace}) 
of grades less than the number of columns $s_1$ of the Young diagram 
$\mathbf{Y}$ characterizing the spin of the particle are differential forms of 
nonzero differential form degree, which implies that they are gauge fields 
associated with the gauge transformations  given by (\ref{unf4}). 
All the fields of grades not less than $s_1$ 
are $0$-forms constituting the so-called Weyl module.  
The lowest-grade field of the Weyl module generalizes the
Weyl tensor of gravity, and is sometimes called the generalized Weyl tensor, 
or primary Weyl tensor. 
The fields of grades higher than $s_1$  can be expressed 
as derivatives of the generalized Weyl tensor via unfolded equations.  
The generalized Weyl tensor or, equivalently, all the $0$-forms 
of the Weyl module given at any point $x_0$,
encode all the perturbative gauge-invariant on-shell degrees of freedom of the system described. 
For a general discussion, see e.g. \cite{Skvortsov:2008sh,Boulanger:2008up,Boulanger:2008kw}.

\section{Metric-like dualisations of gravity}
\label{sec:metric}

Two massless fields in flat spacetime are said to be dual to each other if, on-shell, 
they describe representations of the Wigner little group\footnote{In the present
paper, we consider helicity fields for which the action of the translation 
subalgebra ${t}_{d-2}$ of Wigner's massless little algebra 
${so}(d-2)\subsetplus {t}_{d-2}$ is trivial.} 
${so}(d-2)$ which are related by Hodge dualisation, meaning that
they are actually equivalent.
More precisely, for any massless field propagating in 
Minkowski spacetime and carrying an irreducible representation 
of the Wigner little group given by a ${so}(d-2)$ 
$\mathbf{Y}$-shaped tensor, one can trivially generate, on-shell
and in the light-cone gauge, other equivalent 
irreducible representations by Hodge dualising any (number of) 
columns with totally antisymmetric Levi-Civita rank-$(d-2)$ tensor 
$\epsilon_{i[d-2]}\,$. 
In this paper, we will only dualise fields such that
their corresponding ${so}(d-2)$ representation 
on-shell is described by a tensor of the \emph{same} shape 
as the tensor used for their 
${gl}(d)$ covariant representation off-shell, as 
appearing inside the covariant action. 
\vspace*{.3cm} 

For example \cite{Hull:2001iu}, consider a massless spin-$2$ particle,
on-shell given by the symmetric traceless tensor 
$h_{mn}$ of ${so}(d-2)\,$. 
It can be Hodge dualised to give a traceless ${so}(d-2)$-tensor 
of shape $\mathbf{Y}[d-3,1]$
\begin{equation*}
T^{m[d-3],}{}_{p}=\epsilon^{m[d-3]n}h_{np}\;,
\end{equation*}
which obviously gives an equivalent ${so}(d-2)$-irrep as 
the one corresponding to the original $h_{mn}$ field. 
On the other hand, the representation of ${so}(d-2)$ 
given by $T$ can be uplifted off-shell, in terms of a ${gl}(d)$
field of the same shape, for which the action is known and can be given 
either in the Labastida \cite{Labastida:1987kw} or in the 
Skvortsov \cite{Skvortsov:2008sh} formulation. 
\vspace*{.3cm} 

It is of interest to generate the dual action from the 
action for the original field through the so-called parent action, 
containing fields associated with both equivalent formulations. 
One ends up with one or another dual action, depending on the 
way one eliminates fields through their equations of motion and fixing gauges. 
\vspace*{.3cm} 

One of the ways to write a parent action for a spin-$2$ field
and its dual is as follows \cite{West:2001as,Boulanger:2003vs}.
One starts from the first-order action for linearised gravity, 
formulated in terms of the frame $e_{\bf{1}}^a$ and 
the spin-connection $\omega_{\bf{1}}^{ab}=-\omega_{\bf{1}}^{ab}$,
which is of the form (\ref{action}). Solving $\omega$ in terms 
of $de$ from (\ref{action1}) and plugging it back to 
(\ref{action}) we obtain
\begin{equation}
\label{d1}
S[e_{a|b}]=4\int{d^dx[C_{ca|}{}^aC^{cb|}{}_b-\frac{1}{2}C_{ab|c}C^{ac|b}-
\frac{1}{4}C_{a[2]|c}C^{a[2]|c}]}\;, 
\end{equation}
where $C_{ab|c}=\partial_{[a}e_{b]|c}\,$. The $\lambda$-symmetry with 
$\lambda^{ab}=-\lambda^{ba}$ inherited from (\ref{action}) 
can be used to gauge away the antisymmetric part of $e_{a|b}$, 
so the action (\ref{d1}) depends only on $h_{aa}=e_{(a|a)}$. 
The action (\ref{d1}) is just a rewriting of the linearised action of
general relativity.
\vspace*{.3cm} 

To pass to the parent action we  add one term
\begin{equation}
\label{d2}
S[C_{ab|c},Y_{abc|d}]=4\int{d^dx[-\frac 12 C_{ab|c}\partial_{d}Y^{dab|c}+
C_{ca|}{}^aC^{cb|}{}_b-\frac{1}{2}C_{ab|c}C^{ac|b}-
\frac{1}{4}C_{ab|c}C^{ab|c}]}\;, 
\end{equation}
where $Y_{abc|d}=Y_{[abc]|d}$ and $C_{ab|c}$ is no longer thought as 
a derivative of $e\,$. The field $Y$ can be treated as a Lagrange 
multiplier for the constraint $\partial_{[a}C_{bc]|d}=0$,
which can be solved as $C_{ab|c}=\partial_{[a}e_{b]|c}$ thus 
recovering (\ref{d1}). On the other hand, by examining the equation
of motion for $C$, one sees that $C$ is auxiliary and can 
be eliminated from the action to give
\begin{equation}
\label{d3}
S[Y^{abc|d}]=\int{d^dx[Z_{ab|c}Z^{ac|b}-\frac{1}{d-2}Z_{ab|}{}^{b}
Z^{ac|}{}_c]}\;,
\end{equation}
where $Z^{ab|c}=\partial_dY^{abd|c}\,$. 
It is convenient to rewrite it in terms of the Hodge dual field 
$T^{a[d-3]|c}=\epsilon^{a[d-3]b[3]}Y_{b[3]|}{}^c\,$, which up to 
an overall factor gives
\begin{equation}
S[T^{a[d-3]|b}]=\int d^dx\Big[ X^{a[d-2]|b}X_{a[d-2]|b} 
\label{d4}
-\tfrac{(d-2)^2}{(d-3)}X^{a[d-3]b|}{}_bX_{a[d-3]c|}{}^c+
\tfrac{(d-2)}{(d-3)}X^{a[d-3]b|c}X_{a[d-3]c|b}\Big]\;,
\end{equation}
where $X^{a[d-2]|b}=\partial^{[a}T^{a[d-3]]|b}\,$. 
This action is the analogue of (\ref{d1}) for dual graviton.
Through the parent action it inherits the $\lambda$-symmetry 
of (\ref{d1}). Its Hodge dual $(*\lambda)^{a[d-2]}$ can 
be used to gauge away totally antisymmetric part 
of $T^{a[d-3]|b}$ leading to Labastida's metric-like 
formulation for the massless spin-$\mathbf{Y}[d-3,1]$ field.
\vspace*{.3cm} 

On the other hand, (\ref{d4}) can be recast into the form 
(\ref{action}), where $T^{a[d-3]|b}$ plays role 
of the lowest grade field $e_{\bf{d-3}}^a$. The algebraic symmetry 
$(*\lambda)^{a[d-2]}$ 
signals that the first auxiliary field is a
1-form valued in $\mathbf{Y}[d-2]$-shaped 
tensors, which is exactly the second grade field
(\ref{unfspace}) required by the first order approach (\ref{action}). 
\vspace*{.3cm} 

In the light-cone gauge, on-shell, it is possible to dualise the dual graviton 
so as to produce the double dual graviton described on-shell by 
$\mathbf{Y}[d-3,d-3]$-shaped tensor 
of $so(d-2)$ \cite{Hull:2001iu}
\begin{equation*}
Y^{m[d-3],n[d-3]}=\varepsilon^{m[d-3]p}\varepsilon^{n[d-3]r}h_{pr}\;.
\end{equation*}
The crucial difference with the first dualisation is that $Y$ is no longer 
traceless. Indeed, product of two antisymmetric tensors can be 
rewritten in terms of $\delta$-symbols (\ref{epseps}), so
\begin{equation*}
Y^{m[d-3],}{}_{n[d-3]}=\sigma (d-2)!\delta_{n[d-3]p}^{m[d-3]r}h^{p}{}_{r}=
\sigma (d-2)!\left(\tfrac{1}{d-2}\delta_{n[d-3]}^{m[d-3]}\delta^r_ph^{p}{}_{r}-
\tfrac{(d-3)}{(d-2)}\delta_{[n[d-4]}^{[m[d-4]}h^{m]}{}_{n]}\right)\;.
\end{equation*}
The first term in the bracket vanishes because $h$ is traceless, while 
the second term reveals that $Y$ is what we call 
a pure $(d-4)$-fold trace.\footnote{More generally, in the case a 
tensor $T$ is represented by the direct product
of $m$ metric tensors and another traceless tensor, we say that 
$T$ is  $m$-fold pure-trace.}  
\vspace*{.3cm} 

Such theories can be described  in a ${so}(d-1,1)$-covariant 
way by means of the gauge  invariant curvatures. 
It is easy to show that the Weyl tensor of gravity being 
a traceless $so(d-1,1)$-tensor of shape $\mathbf{Y}[2,2]\,$,
after double dualisation it gives rise to a ${gl}(d)$ tensor
$\widetilde{C}$ of shape $\mathbf{Y}[d-2,d-2]\,$, formed from the 
tensor product of $(d-4)$ metric tensors and a traceless
$\mathbf{Y}[2,2]\,$ tensor, \emph{i.e.} $\widetilde{C}$ is 
$(d-4)$-fold pure-trace. The traceless $\mathbf{Y}[2,2]\,$-tensor
appearing in the representation of $\widetilde{C}$ is of course
the Weyl tensor of linearised gravity. 
The tensor $\widetilde{C}$ can be identified with the Weyl tensor of a 
double dual graviton. 
Imposing on the curvature of the double-dual graviton the higher-power
tracelessness equations written in \cite{Hull:2000zn}, 
one can describe proper degrees of freedom. 
Let us recall here 
the results of the section 6.3. of \cite{Bekaert:2002dt} to which we refer 
for more details.
In the case of an irreducible ${gl}(d)$ gauge field 
$\varphi_s$ with $s$ columns, one makes the following assumption 
concerning the positive integers 
$l_i$ $(i=1,\ldots,s)$ associated with the multiform curvature 
$K=d^s\varphi\in\Omega_{[s]}^{l_1,\ldots,l_s}$ inside the 
differential complex where $d^{s+1}\equiv 0\,$: 
\begin{equation}
l_i+l_j\leqslant d\,,\quad
\forall\, i,j\; .
\label{hypoth} 
\end{equation} 
In that case the local, covariant field equations read 
\begin{equation} 
\mbox{Tr}_{ij}\{K\}=0\,,\quad \forall \, i,j\;.
\label{FIeld} 
\end{equation}
Denoting by $*_i$ the Hodge duality operation 
acting in the $i$th column of a 
multiform, the field equations (\ref{FIeld}) combined with the 
algebraic Bianchi identities 
$\mbox{Tr}_{ij}\{\,*_i\,K\}=0\,,\; \forall i,j:\,\, 1\leqslant i<j\leqslant s\,$  
state that the curvature $K$ is a tensor irreducible under 
the orthogonal group.
To any non-empty subset $I\subset\{1,2, \dots, s\}$ ($\#I=m$), one
associates a Hodge duality operator
$*_I :=  \prod_{k \in I}*_k \,$. 
The dual $*_I K$ of the curvature is a multiform in
$\Omega^{\ell_1,\ldots,\ell_s}_{[s]}\,$, 
where the lengths $\ell_i$ are defined by
\begin{equation}
\ell_i \equiv \left\{
\begin{array}{lll}l_i\,\quad&\mbox{if}\,\,i\not\in I\,,&
\\
d-l_i\,\quad&\mbox{if}\,\,i\in I\;.& 
\end{array}\right.
\label{ell}
\end{equation}
\vspace*{.1cm}
 
It can be proved \cite{Bekaert:2002dt} that 
the algebraic Bianchi identities together with the field
equations (\ref{FIeld}) imply the relations 
\begin{equation}
\mbox{Tr}_{ij}\,\{*_i\,(*_I K\,)\}\,=\,0\,,\quad \forall\; i,j
\; :\; 
\ell_j\leqslant \ell_i\quad, 
\label{from}
\end{equation}
where $\ell_i$ is the length (\ref{ell}) of the $i$th column of $*_I K\,$. 
One defines
$\widetilde{K}_I$ to be the multiform obtained after 
reordering the columns of $*_I K\,$, such that the heights of
the columns of $\widetilde{K}_I$ are non-increasing.  
The identity (\ref{from}) can then be formulated as
${\mbox{Tr}}_{ij}\{*_i \widetilde{K}_I\}=0\,,~\forall
 ~i,j~:~1\leqslant i<j\leqslant s\,$, implying that 
$\widetilde{K}_I$ is ${gl}(d)$-irreducible. 
The differential Bianchi identities $d_iK=0$ together with the
field equations (\ref{FIeld}) imply that $d_i*_iK=0\,$, which entails 
$d_i(*_I{K})=0\,,\quad \forall i\in\{1,\ldots,s\}\,$. 
As a result of the generalized Poincar\'e Lemma 
proved in \cite{Bekaert:2002dt}, one has
$\widetilde{K}_I = d_1 d_2\ldots d_s \tilde{\kappa}_I$
for some gauge field 
$\tilde{\kappa}_I\,$. 
The Hodge operators $*_I$ therefore relate different free field
theories of arbitrary tensor gauge fields, extending the
electric-magnetic duality property of electrodynamics. 
In the same way, one obtains the field equations of the dual theory
\be 
\mbox{Tr}^{m_{ij}}_{ij}\,\{\,*_I K\,\}\,=\,0\,,\quad \forall ~
i,j:\,i<j\,
\label{genfieldequ}
\ee 
where 
\be
m_{ij}\equiv
\left\{\begin{array}{lll}1+D-l_i-l_j\quad&\mbox{if}\,\,i\,\,\mbox{and}\,\,j\in
I\,,&\\
1\,\quad&\mbox{if}\,\,i\,\,\mbox{or}\,\,j\not\in
I\,.&\end{array}\right.
\label{mij}
\ee
\vspace*{.5cm}

On-shell formulations involving higher powers of the trace operation 
generically are not Lagrangian, and the double-dual spin-2 case is 
the paradigmatic example discussed in \cite{Hull:2000zn}.
In order to set up the ${gl}(d)$-covariant formalism 
needed for the description of propagating gauge fields that
become pure ${so}(d-2)$-trace on-shell, some work has to be done 
since we explained that the Labastida--Skvortsov representation for the 
covariant field is not suitable in those cases.
In a previous work with P. P. Cook \cite{Boulanger:2012df}, 
we have given a metric-like action for the double-dual
graviton, and the purpose of the present paper is to give 
a frame-like action that allows to treat the arbitrary 
mixed-symmetry cases as well. 
\vspace*{.5cm}

Staying at the on-shell level for the moment, 
one can use the unfolded equations of motion (\ref{unfmix})
 with the properly  modified trace constraints
in order to describe double dual linearised 
gravity. 
As we mentioned above, for a propagating 
spin-$\mathbf{Y}[d-3,d-3]$ gauge field in flat spacetime of 
dimension $d\,$, 
the shapes of the Weyl-module tensors given by (\ref{unfspace}) are related 
(by the double Hodge dualisation with rank-$d$
antisymmetric tensor in fiber space) to the Weyl-module tensors 
of gravity.  
In order  for the two Weyl modules to describe the same degrees 
of freedom, their trace constraints should also be related by the same
fiber-space Hodge dualisation\footnote{Let us note that if we allow 
$\sigma_-$ to contain Levi-Civita 
symbols we can always replace any field 
of the unfolded system by its dual 
if this is accompanied by the proper $\sigma_-$ redefinition. So, one can make 
Weyl modules of dual theories exactly coinciding.} in a way that 
generalises the equations (\ref{genfieldequ}) and (\ref{mij}). 
Analogously to the 
previous discussion, it implies that the Weyl-module 
tensors for the double dual graviton should be $(d-4)$-fold pure-trace.
The question is what gauge potentials 
should be used to make the theory Lagrangian. 
One can see that gauge potentials (\ref{unfspace}) do not work because  
by construction the lower grade field describes traceless 
(not $(d-4)$-fold pure-trace as it should be) $so(d-2)$
$\mathbf{Y}[d-3,d-3]$-shaped tensor after all the gauge degrees of 
freedom are factored out.
 
\section{Frame-like dualisations of linearised gravity}
\label{fldg}

In the previous Section we have shown how the metric-like theories can be 
dualised through the concept of the parent action. It appears that the same 
can be done in a more economic way purely in terms of frame-like 
fields and frame-like actions. In this Section we will illustrate the
frame-like dualisation in all details 
for the case of the first and the second dualisations of gravity.

\subsection{The dual gravity}

In the particular case of linearised gravity the spin is $\mathbf{Y}[1,1]$ and 
the first two equations of (\ref{unfmix}) acquire the form
\begin{eqnarray}
\label{eq1}
T_{\bf{2}}^a &:=& de_{\bf{1}}^a+h_{b}\omega_{\bf{1}}^{ab}=0\;,
\\
\label{eq2}
R_{\bf{2}}^{a[2]} &:=& d\omega_{\bf{1}}^{a[2]}+
h_{b}h_{b}C_{\bf{0}}^{a[2],b[2]}=0\;.
\end{eqnarray}
They come from the action (\ref{action}) 
\begin{equation}
\label{acgr}
S=\int{(de^a_{\bf{1}}+\frac{1}{2}h_{b}\omega_{\bf{1}}^{ab})
\omega_{\bf{1}}^{cd}H_{acd}}\;.
\end{equation}
\vspace*{.3cm}

The parent action for the first dualisation is
\begin{equation}
\label{acgr1}
S=\int{\left[(de^a_{\bf{1}}+\frac{1}{2}h_{b}\omega_{\bf{1}}^{ab}+
t_{\bf{2}}^a)
\omega_{\bf{1}}^{cd}H_{acd}+t_{{\bf{2}}a}d\,{\tilde{e}}_{{\bf d-3}}^{\,a}\right]} \;.
\end{equation}
Here $t_{\bf{2}}^a$ is a torsion-like auxiliary field and ${\tilde{e}}_{{\bf d-3}}^{\,a}$
will be identified with the frame-field for the dual graviton. 
This action is invariant under the following gauge transformations
\begin{eqnarray}
 \label{cl14}
 \delta e_{\bf{1}}^a &=& d\xi_{\bf{0}}^a+h_{b}\lambda_{\bf{0}}^{ab}
 -\psi_{\bf{1}}^a,\quad \delta \xi_{\bf{0}}^a = \bar\psi_{\bf{0}}^{a}\;,
\\
 \label{cl15}
 \delta \omega_{\bf{1}}^{a[2]} &=& d\lambda_{\bf{0}}^{a[2]}\;,
\\
 \label{cl16}
 \delta t_{\bf{2}}^a &=& d\psi_{\bf{1}}^a,\quad \delta \psi_{\bf{1}}^a = d \bar\psi_{\bf{0}}^{a}\;,
\\
 \label{cl17}
 \delta\tilde e_{{\bf d-3}}^a &=& d \tilde \xi_{{\bf d-4}}^a-
 h_{l[d-3]}(*\lambda_{\bf{0}})^{al[d-3]},\quad \delta \tilde \xi_{{\bf d-4}}^a = d\bar{\tilde \xi}{}_{{\bf d-5}}^a
\;,\quad \ldots
 \end{eqnarray}
\vspace*{.1cm}

In order to show that the action 
(\ref{acgr1}) is equivalent to the original action (\ref{acgr}), one should 
treat ${\tilde{e}}_{{\bf d-3}}^{\,a}$ as a Lagrange multiplier for the constraint 
$dt_{\bf{2}}^a=0\,$,
which can be solved as $t_{\bf{2}}^a=d\beta_{\bf{1}}^a\,$. 
Then $t_{\bf{2}}^a$ can be set to zero by performing a gauge transformation (\ref{cl16})
with appropriate parameter $\psi_{\bf{1}}^a\,$, 
leading to the linearised gravity action (\ref{acgr}).
\vspace*{.3cm}

The field equations derived from the action (\ref{acgr1}) can be promoted to the following unfolded form
\begin{eqnarray}
 \label{cl9}
 T_{\bf{2}}^a &:=& de_{\bf{1}}^a+h_{b}\omega_{\bf{1}}^{ab} + t_{\bf{2}}^a = 0\;,
\\
 \label{cl10}
 R_{\bf{2}}^{a[2]} &:=&  d\omega_{\bf{1}}^{a[2]} + h_{b}h_{b}C_{\bf{0}}^{a[2],b[2]} = 0\;,
\\
 \label{cl11}
 \tau_{\bf{3}}^a &:=& dt_{\bf{2}}^a = 0 \;,
\\
 \label{cl12}
 \tilde T_{\bf{d-2}}^a &:=& d\tilde e_{{\bf d-3}}^a+(-1)^{d-3} h_{l[d-3]}(*\omega)_{\bf{1}}^{al[d-3]} = 0\;,
\end{eqnarray}
where 
\begin{equation}
\label{cl13}
(*\omega)_{\bf{1}}^{a[d-2]}:=\varepsilon^{a[d-2]b[2]}\omega_{{\bf{1}} b[2]}\;.
\end{equation}
The corresponding unfolded equations for the zero-forms 
are identical to those for linearised gravity~\cite{Vasiliev:2001wa}:
\begin{equation}
 d C_{\bf{0}}^{a(k),b(2)} + e_c\,C_{\bf{0}}^{c\{a(k),b(2)\} } ~=~ 0\quad, \quad k=2,3,\ldots
\end{equation} 
where the curly brackets ``$\{ \}$'' denote projection on the symmetry of the
tensor appearing under the differential. 
The equations (\ref{cl9})-(\ref{cl13}) are manifestly gauge invariant with respect to
the gauge transformations (\ref{cl14})-(\ref{cl17}), where the zero-forms do not
transform.  
\vspace*{.3cm}

The algebraic $\psi$-symmetry in (\ref{cl14})
can be used to gauge away the frame field $e^a_{\bf{1}}\,$. 
The gauge transformation for the gauge parameter $\xi_{\bf{0}}^a\,$, \emph{viz.}
$\delta \xi_{\bf{0}}^a=\bar \psi_{\bf{0}}^a\,$,
implies that the $\xi^a_{\bf{0}}$ gauge parameter can be shifted to zero.
The equation  (\ref{cl9}) can be used to express $t$ in terms of $\omega$
\begin{equation*}
t_{\bf{2}}^a=-h_{b}\omega_{\bf{1}}^{ab}\;.
\end{equation*} 
Substituting this back to (\ref{acgr1}) gives
\begin{equation}
\label{cl18}
S=\frac{(-1)^d}{2!(d-2)!\sigma}\int{\left(d\tilde e^a_{\bf d-3}+\frac 12(-1)^{d-3} 
h_{b[d-3]}(*\omega)_{\bf 1}^{ab[d-3]}\right)(*\omega)_{\bf 1}^{c[d-2]}H_{ac[d-2]}}\;.
\end{equation}
Identifying $(-1)^{d-3}(*\omega)_{\bf 1}^{ab[d-3]}$ as a connection 
$\tilde\omega$ of the dual theory we recover the lower grade fields and the action 
of the frame-like formulation for $\mathbf{Y}[d-3,1]$-spin field.

\subsection{The double-dual graviton}

The second dualisation can be performed analogously.
We start from the frame-like action of dual gravity\footnote{Note that  
we reset the notation, so as to avoid cumbersome ``double-tilde" 
$\tilde{\tilde{e}}$ notation.}
\begin{equation}
 \label{cl19}
 S=\int{\left(d e^a_{\bf d-3}+\frac 12 
 h_{b[d-3]}\omega_{\bf 1}^{ab[d-3]}\right)\omega_{\bf 1}^{c[d-2]}H_{ac[d-2]}}\;.
 \end{equation}
Following the strategy of the first dualisation we should introduce an 
auxiliary field $t$ such that its differential symmetry (\ref{unf5}) 
acts on the lowest grade field $e$ of the original theory in an algebraic 
way and can be used to gauge it away. 
Then we should introduce a new frame field $\tilde e$
such that $td\tilde e$ is a $d$-form scalar.  
\vspace*{.3cm}
 
The first option is to take $t$ to be a $(d-2)$-form with one fiber index 
$t^a_{\bf d-2}\,$.
The associated gauge parameter $\psi^a_{\bf d-3}$ has the same form degree 
as the form degree of the frame field $e^a_{\bf d-3}$ and takes its 
values in the same representation space, 
so it is appropriate to gauge $e^a_{\bf d-3}$ away. 
The parent action is
\begin{equation}
\label{cl20}
S=\int{\Big[(d e^a_{\bf d-3}+\frac 12 
h_{b[d-3]}\omega_{\bf 1}^{ab[d-3]}+t^a_{\bf d-2})\omega_{\bf 1}^{c[d-2]}H_{ac[d-2]}+
t^a_{\bf d-2}d\tilde e_{\bf 1}^a\Big]}\;.
\end{equation}
It is easy to see that it dualises the dual graviton back to the usual Fierz--Pauli graviton.
\vspace*{.3cm}

To make the second non-trivial dualisation one should choose the auxiliary 
field $t$ to be a $2$-form valued transforming like a $\mathbf{Y}[d-3]$-
type Lorentz tensors. Its differential 
gauge parameter  $\psi^{a[d-3]}_{\bf 1}$ contains enough degrees of 
freedom to gauge away the frame field completely. 
The corresponding parent action is now
\begin{equation*}
S[e^a_{\bf d-3},\omega_{\bf 1}^{ab[d-3]},
t_{\bf 2}^{a[d-3]},\tilde e_{\bf d-3}^{a[d-3]}]=\int{\Big[(de^a_{\bf d-3}
+\frac 12 h_{b[d-3]}\omega_{\bf 1}^{ab[d-3]}+
 h_{b[d-4]}t_{\bf 2}^{ab[d-4]})\omega_{\bf 1}^{c[d-2]}H_{ac[d-2]}}
\end{equation*}
\begin{equation}
\label{00frd9}
 +(-1)^{d-1} t_{{\bf 2}a[d-3]}d\tilde e_{\bf d-3}^{a[d-3]}+
(-1)^{d-1}\frac{\alpha}{2} t_{{\bf 2}a[d-3]}h^{a[d-5]}h_c
(*t)_{\bf 2}^{a[2]c}\Big]\;,
\end{equation}
where 
\begin{equation*}
(*\omega)^{a[2]}_{\bf 1}=\varepsilon^{a[2]b[d-2]}\omega_{{\bf 1}b[d-2]}, 
\quad 
(*t)^{a[3]}_{\bf 2}=\varepsilon^{a[3]b[d-3]}t_{{\bf 2}b[d-3]}\;.
\end{equation*}
Notice the last term bilinear in the auxiliary field $t$, which has no analogue in 
the parent action (\ref{acgr1})\footnote{It cannot be constructed
there because of the form degree and Lorentz symmetries of 
$t_{\bf 2}^a\,$.}.
The coefficient $\alpha$ is arbitrary although it has two special values
which will briefly be discussed later.
\vspace*{.3cm}

The manifest gauge symmetries of the above action read
\begin{eqnarray}
\label{ddg5}
\delta e_{\bf{d-3}}^a &=& d\xi_{\bf{d-4}}^a+(-1)^{d-2}h_{b[d-3]}\lambda_{\bf{0}}^{ab[d-3]}
+(-1)^{d-3}h_{b[d-4]}\psi_{\bf{1}}^{ab[d-4]}\;,
\\
\label{ddg6}
\delta\omega_{\bf{1}}^{a[d-2]} &=&
d\lambda_{\bf{0}}^{a[d-2]}\;,
\\
\label{ddg7}
\delta t_{\bf{2}}^{a[d-3]} &=& d\psi_{\bf{1}}^{a[d-3]}\;,
\\
\label{ddg8}
\delta \tilde e_{{\bf d-3}}^{a[d-3]}  &=& 
d\tilde \xi_{{\bf d-4}}^{a[d-3]}+
(-1)^{d-3}h^{[a[d-4]}h_b(*\lambda){}_{\bf{0}}^{a]b}+(-1)^{d-4}
\alpha h^{[a[d-5]}h_b(*\psi)_{\bf 1}^{a[2]]b}\;.
\end{eqnarray}

To show the equivalence with dual gravity one should treat 
$\tilde e_{\bf d-3}^{a[d-3]}$ as a Lagrange multiplier for the constraint 
$dt_{\bf 2}^{a[d-3]}=0\,$, which entails
$t_{\bf 2}^{a[d-3]}=d\beta_{\bf 1}^{a[d-3]}\,$.
Using the gauge symmetry (\ref{ddg7}), one can set $t_{\bf 2}^{a[d-3]}$
to zero and recuperate the frame-like formulation for linearised dual 
gravity. 
\vspace*{.3cm}
 
On the other hand, the fields equations for (\ref{00frd9}) 
can be promoted to their unfolded form:
\begin{eqnarray}
\label{ddg1}
T_{\bf{d-2}}^a &:=& de_{\bf{d-3}}^a+h_{b[d-3]}\omega_{\bf{1}}^{ab[d-3]}
+h_{b[d-4]}t_{\bf{2}}^{ab[d-4]}=0\;,
\\
\label{ddg2}
R_{\bf{2}}^{a[d-2]} &:=&
d\omega_{\bf{1}}^{a[d-2]}+
h_{b}h_{b}C_{\bf{0}}^{a[d-2],b[2]}=0\;,
\\
\label{ddg3}
\tau_{\bf{3}}^{a[d-3]} &:=& dt_{\bf{2}}^{a[d-3]}=0\;,
\\
\label{ddg4}
\widetilde T_{\bf{d-2}}^{a[d-3]} &:=& d\tilde e_{{\bf d-3}}^{a[d-3]}+
h^{a[d-4]}h_b(*\omega){}_{\bf{1}}^{ab}+\alpha h^{a[d-5]}h_b(*t)_{\bf 2}^{a[2]b}=0\;.
\end{eqnarray}
Complemented  with the unfolded equations for the Weyl module of massless 
spin-$\mathbf{Y}[d-3,1]\,$, the resulting set of equations
satisfy the compatibility conditions (\ref{unf3}). 
Taking into account both fiber and form indices,  $e_{\bf{d-3}}^a$
 and $\psi_{\bf{1}}^{a[d-3]}$ carry the representation of the Lorentz 
group given by the outer product of $\mathbf{Y}[1]$ and $\mathbf{Y}[d-3]$, 
which can be 
decomposed into irreducible parts\footnote{The multiplication rules for 
$GL(d)$ and 
$SO(m,n)$ representations are given, for example, in 
\cite{Barut:1986dd, FulHar, Bekaert:2006py}}
\begin{equation}
\mathbf{Y}[1] \otimes \mathbf{Y}[d-3]= \mathbf{Y}[d-2]\oplus \mathbf{Y}[d-3,1]\oplus
\mathbf{Y}[d-4]\;.
\end{equation}
According to this decomposition, $\psi_{\bf{1}}^{a[d-3]}$ can be presented in the 
form
\begin{equation*}
\psi_{\bf{1}}^{a[d-3]}=h_b\phi_{\bf{0}}^{a[d-3]b}+h_b\phi_{\bf{0}}^{a[d-3],b}+
h^{a}\phi_{\bf{0}}^{a[d-4]}\;,
\end{equation*}
each term being irreducible.
It is easy to check that none of these terms are annihilated 
by the operator in front of $\psi_{\bf{1}}^{a[d-3]}$ in (\ref{ddg5}), so 
$e_{\bf{d-3}}^a$ can be completely gauged away by the $\psi$-symmetry.
\vspace*{.3cm}

An important difference with the first dualisation case is that, 
in spite of the fact that the frame field $e_{\bf{d-3}}^{a}$ of the 
original theory can be completely gauged away, some gauge 
parameters associated with $e_{\bf{d-3}}^{a}$ survive and play an
important role, as we  explain now:
The gauge  transformations for the gauge parameter $\xi_{\bf{d-4}}^a$ are
\begin{equation}
 \label{ddg9}
 \delta \xi_{\bf{d-4}}^a=d\bar\xi_{\bf{d-5}}^a+h_{b[d-4]}
 \bar\psi_{\bf{0}}^{ab[d-4]}\,,
 \end{equation}
where $\bar\psi_{\bf{0}}^{a[d-3]}$ is the second-level gauge parameter 
associated with $t_{\bf{2}}^{a[d-3]}$.
The gauge parameter $\xi_{\bf{d-4}}^a$ decomposes into the following
irreducible representations of the Lorentz group
\begin{equation*}
\mathbf{Y}[1] \otimes \mathbf{Y}[d-4] \cong \mathbf{Y}[d-3]\oplus \mathbf{Y}
[d-4,1]\oplus \mathbf{Y}[d-5]\;.
\end{equation*}
Only the $\mathbf{Y}[d-3]$ component
can be gauged away by the algebraic second order $\bar\psi$ symmetry (\ref{ddg9}). 
\vspace*{.3cm}

The remaining components of $\xi_{\bf{d-4}}^a$ transform $e_{\bf{d-3}}^a$ 
as in (\ref{ddg5}) and should be accompanied by the 
proper algebraic $\psi$-shifts $\psi(\xi)\sim d\xi$ in order 
to preserve the gauge $e_{\bf{d-3}}^a=0\,$. 
These compensating $\psi(\xi)$ transformations act on the new frame 
field $\tilde{e}_{\bf{d-3}}^{a[d-3]}$ because of the 
last term in (\ref{ddg8}), thus leading to differential gauge 
transformations containing divergences of the gauge parameter, 
and not only curls as for usual mixed-symmetry fields
traceless on-shell.
\vspace*{.3cm}

Another difference between the second and the first dualisations 
is that the auxiliary field $t_{\bf{2}}^{a[d-3]}$ 
cannot be fully expressed in terms of 
$\omega_{\bf{1}}^{a[d-2]}\,$. Indeed, it is 
easy to see that some irreducible components of $t_{\bf{2}}^{a[d-3]}$ 
are annihilated in (\ref{ddg1}).
So, the action (\ref{00frd9}) cannot be further simplified
to the first-order form where only the frame-like field 
$\tilde{e}_{\bf{d-3}}^{a[d-3]}$ of double-dual gravity
and the connection one-form $\omega_{\bf{1}}^{a[d-2]}\,$ 
would enter. Actually, it is possible to trade 
$\omega_{\bf{1}}^{a[d-2]}\,$ for the connection 
$\omega_{\bf{d-3}}^{a[d-2]}\,$ associated with the 
 frame field $\tilde{e}_{\bf{d-3}}^{a[d-3]}$ in the approach of \cite{Skvortsov:2008vs} , 
by following the same Hodge dualisation steps that we used in order
to go from the frame fields ${e}_{\bf{d-3}}^{a}$ to    
$\tilde{e}_{\bf{d-3}}^{a[d-3]}\,$. 
\vspace*{.3cm}
 
The auxiliary field $t_{\bf{2}}^{a[d-3]}$ plays a role of a supplementary
connection for $\tilde{e}_{\bf{d-3}}^{a[d-3]}\,$: 
Some components of it can be expressed in terms of one derivative of 
$\tilde{e}_{\bf{d-3}}^{a[d-3]}\,$ via a projection of (\ref{ddg4}) 
which annihilates $\omega_{\bf{1}}^{a[d-2]}\,$.
Then, plugging the result in (\ref{ddg3}) gives 
second-order dynamical equations for $\tilde{e}_{\bf{d-3}}^{a[d-3]}\,$ 
additional to those given by (\ref{ddg2}).
\vspace*{.3cm}

Although the parent action and unfolding principles guarantee that the 
action (\ref{00frd9}) can indeed be used to describe linearised gravity 
by means of the double dual field $\tilde e_{\bf d-3}^{a[d-3]}\,$,
it is instructive to show it more explicitly by a direct 
counting of degrees of freedom. 
First, one should exhibit the dynamical fields and differential gauge 
parameters of the unfolded system (\ref{ddg1})-(\ref{ddg4}). 
Then one can perform a counting of degrees of counting as follows. 
(For simplicity we consider here the $d=5$ case.) 
\vspace*{.3cm}

The full set of fields and gauge 
parameters is
\begin{eqnarray}
e_{\bf{2}}^a\quad &\rightarrow &  \quad \xi_{\bf{1}}^a
\quad \rightarrow \quad \bar\xi_{\bf{0}}^a\;,
\nonumber \\
 \omega_{\bf{1}}^{a[3]}\quad &\rightarrow& \quad
\lambda_{\bf{0}}^{a[3]}\;,
\nonumber \\
 t_{\bf{2}}^{a[2]}\quad &\rightarrow & \quad\psi_{\bf{1}}^{a[2]}
 \quad \rightarrow \quad \bar\psi_{\bf{0}}^{a[2]}\;,
 \nonumber \\
 \label{ddsf}
 \tilde e_{{\bf 2}}^{a[2]} \quad &\rightarrow &\quad \tilde \xi_{{\bf 1}}^{a[2]}
 \quad \rightarrow \quad \bar{\tilde \xi}_{{\bf 0}}^{a[2]}\;.
\end{eqnarray}
The dynamical field is given by  $\tilde e_{{\bf 2}}^{a[2]}$ modulo
pure gauge shifts $\lambda_{\bf{0}}^{a[3]}\,$, see (\ref{ddg8}). 
As a result 
(we use Hodge dualisations in order to simplify the Young diagrams)
\begin{equation}
\label{dddf}
\text{dyn. field:} \quad (\mathbf{Y}[2]\otimes\mathbf{Y}[2]) 
\ominus\mathbf{Y}[3] \cong
\mathbf{Y}[2,2]\oplus \mathbf{Y}[2,1]\oplus \mathbf{Y}[1,1]\oplus\mathbf{Y}[1]
\oplus \mathbf{Y}[0]\;.
\end{equation}
By decomposing the zero-torsion-like equations (\ref{ddg1}) and 
(\ref{ddg4}) into irreducible Lorentz components, it is straightforward 
to see that the fields 
$t_{\bf{2}}^{a[2]}$ and $\omega_{\bf{1}}^{a[3]}$ are auxiliary, 
being fully expressed in terms of the derivative of 
$\tilde e_{\bf{2}}^{a[2]}$ and $e_{\bf{2}}^{a}\,$, 
the latter being pure gauge.
\vspace*{.3cm}

\noindent The first-order differential gauge parameters are given by $\tilde\xi_{{\bf 1}}^{a[2]}$  
plus $\xi_{\bf{1}}^a$ modulo redundancy $\bar\psi_{\bf{0}}^{a[2]}$. Altogether it 
yields
\begin{equation}
\label{dddf1}
\text{diff. 1st:} \quad (\mathbf{Y}[2]\otimes\mathbf{Y}[1])
\oplus \left[ (\mathbf{Y}[1]\otimes\mathbf{Y}[1]) \ominus \mathbf{Y}[2]\right]\cong 
\mathbf{Y}[2,1]\oplus \mathbf{Y}[2]\oplus \mathbf{Y}[1,1]\oplus\mathbf{Y}[1]
\oplus \mathbf{Y}[0]\;.
\end{equation}
The second-order gauge parameters are $\bar\xi_{\bf{0}}^a$ and 
$\bar{\tilde \xi}_{{\bf 0}}^{a[2]}$ and decompose as follows
\begin{equation}
\label{dddf2}
\text{diff. 2nd:} \quad \mathbf{Y}[2]
\oplus \mathbf{Y}[1]\;.
\end{equation}
This reproduces the set of fields and gauge symmetries found in 
\cite{Boulanger:2012df}. 

Let us stress that once the dynamical fields and differential gauge 
symmetries (\ref{dddf})-(\ref{dddf1}) are known, 
by taking advantage of the property of unfolding that all the gauge 
symmetries are manifest (that is associated 
to fields according to (\ref{unf5})), it is then possible to recover 
unambiguously the complete set of frame-like fields (\ref{ddsf}) 
entering the unfolded formulation. 
In other words, knowing the fields and full set of gauge parameters
entering the metric formulation in \cite{Boulanger:2012df}, one can 
build the spectrum (\ref{ddsf}).
This set of fields gives a hint how to construct the frame-like 
parent action (\ref{00frd9}) for the second dualisation.
Although the auxiliary field $t_{\bf{2}}^a$ is not required for the 
unfolding of dual gravity, see \cite{Skvortsov:2008vs}, 
the pattern of auxiliary fields for the second dualisation of linearised 
gravity can be directly generalised to the first dualisation case, 
thereby providing the parent action (\ref{acgr1}). 
\vspace*{.3cm}

By looking at the Weyl module entering the equations (\ref{ddg2}), 
the principle of unfolding guarantees that the propagating
degrees of freedom (contained in the zero-form module) precisely
correspond to those of linearised gravity via the dual 
$C_{\bf 0}^{a[2],b[2]}\,$ of the primary
Weyl tensor $C_{\bf 0}^{a[d-2],b[2]}\,$ of dual gravity. 
It is nevertheless 
instructive to reproduce this counting using explicitly 
the $p$-form modules with $p>0$ and the experience acquired from the 
Hamiltonian 
analysis of constrained systems \cite{Henneaux:1992ig}.
The fields, first-order and second-order differential gauge 
parameters contribute to the counting of degrees of freedom 
with the multiplicities $1\,$, $2$ and $3$ respectively:
\begin{equation}
\label{degoffree}
SO(4,1): \qquad \left\{
\begin{array}{c}
1\times (\mathbf{Y}[2,2]\oplus \mathbf{Y}[2,1]\oplus \mathbf{Y}[1,1]\oplus\mathbf{Y}[1]
\oplus \mathbf{Y}[0])\\
\\
 2\times (\mathbf{Y}[2,1]\oplus \mathbf{Y}[2]\oplus \mathbf{Y}[1,1]\oplus\mathbf{Y}[1]
\oplus \mathbf{Y}[0])\\
\\
3\times (\mathbf{Y}[2]
\oplus \mathbf{Y}[1])\;.\\
\end{array}
\right.
\end{equation}
Then, continuing with this heuristic procedure, 
one performs dimensional reduction of $SO(4,1)$ tensors to $SO(4)$
and make pairwise cancellations between two adjacent levels, 
thereby obtaining
\begin{equation}
\label{degoffree1}
SO(4): \qquad \left\{
\begin{array}{c}
\mathbf{Y}[2,2]\oplus\mathbf{Y}[1]\\
\\
 \mathbf{Y}[2,1]\oplus  \mathbf{Y}[1,1]\oplus\mathbf{Y}[0]\\
\\
\mathbf{Y}[2]
\oplus \mathbf{Y}[1]\;.\\
\end{array}
\right.
\end{equation}
A dimensional reduction to the Wigner little group $SO(3)$ finally gives 
$\mathbf{Y}[2,1]\,$, 
which is equivalent to the graviational spin-$\mathbf{Y}[1,1]$ field in 
$d=5\,$. 
\vspace*{.3cm}

Let us now find the dynamical equations. 
As we already mentioned, Eq. (\ref{ddg4})  
$ d\tilde e_{{\bf d-3}}^{a[d-3]}+
 h^{[a[d-4]}h_b(*\omega_{\bf{1}})^{a]b}+\alpha h^{[a[d-5]}h_b(*t)_{\bf 2}^{a[2]]b}=0\,$ is a constraint
in the sense that it expresses $\omega_{\bf{1}}^{a[d-2]}$ or $t_{\bf{2}}^{a[d-3]}$ 
in terms of derivatives of $\tilde e\,$. 
In $d=5$  (\ref{ddg2}) carries $\mathbf{Y}[2]\otimes \mathbf{Y}[3]$ 
representation of the Lorentz group. 
Its projection to the representation $\mathbf{Y}[3,2]$ of the Lorentz 
group is a constraint because it expresses the primary Weyl tensor $C$ 
in terms of the lower grade fields. This equations gives the 
gluing of the $p$-form module to the zero-form, Weyl module. 
In addition, the curvature $R$ contributes to the Bianchi identity
\begin{equation*}
d\tilde T_{{\bf 3}}^{a[2]}+
h^{[a}h_b(*R){}_{\bf{2}}^{a]b}-\alpha h_b(*\tau)_{\bf 3}^{a[2]b}=0\;,
\end{equation*}
carrying $\mathbf{Y}[4]\otimes\mathbf{Y}[2]$ representation of the Lorentz group, which
means that all the projections of $R=0$ to $\mathbf{Y}[4]\otimes\mathbf{Y}[2]$ 
are consequences of other equations appearing at the lower grade. 
As a result,
the dynamical part of the $R=0$ is a projection to 
\begin{equation}
\label{ddgravd1}
(\mathbf{Y}[2]\otimes \mathbf{Y}[3])\ominus \mathbf{Y}[3,2]
\ominus(\mathbf{Y}[4]\otimes\mathbf{Y}[2])=
\mathbf{Y}[1,1]\oplus\mathbf{Y}[0]
\end{equation} 
similarly to the ordinary gravity.
\vspace*{.3cm}

As we discussed previously, the equation (\ref{ddg3}) also imposes 
dynamical equations on $\tilde{e}\,$. It carries the 
$\mathbf{Y}[3]\otimes \mathbf{Y}[2]$ tensor representation 
of the Lorentz group. The remaining 
Bianchi identity
\begin{equation*}
 dT_{\bf{3}}^a-h_{b[2]}R_{\bf{2}}^{ab[2]}
 +h_{b}\tau_{\bf{3}}^{ab}=0
 \end{equation*}
implies that projections of $\tau=0$ to $\mathbf{Y}[4]\otimes\mathbf{Y}[1]$ are 
consequences of other equations that have already been taken into account. 
So, in addition it yields second-order dynamical equations on 
$\tilde{e}$, which take their values in the following
representations of the Lorentz group
\begin{equation}
\label{ddgravd2}
(\mathbf{Y}[3]\otimes \mathbf{Y}[2])\ominus(\mathbf{Y}[4]\otimes\mathbf{Y}[1])=
\mathbf{Y}[2,2]\oplus\mathbf{Y}[2,1]\oplus \mathbf{Y}[1]
\;.
\end{equation} 

The equations (\ref{ddgravd1}) and (\ref{ddgravd2}) transform
in the same representation as the field $\tilde{e}\,$, 
see equation (\ref{dddf}), and give equations whose left-hand side
contains the D'Alembertian of $\tilde{e}\,$ plus other second-order
derivative terms that ensure gauge-invariance of the equations. 
This was to be expected from equations derived from an action. 
\vspace*{.3cm}

Although the action (\ref{00frd9}) cannot be simplified in such 
a way as to remain frame-like and at the same time with all the fields
of the dual-graviton sector eliminated, 
it is possible to formulate the action in a metric-like way 
in terms of the dynamical fields valued in the representations 
(\ref{dddf}).
The independent derivation of this metric-like action 
was done explicitly in \cite{Boulanger:2012df}.
\vspace*{.3cm}

There are two special values for $\alpha$. When $\alpha=0$
the last term in (\ref{00frd9}) vanishes. It implies that
the last term on the right-hand side of (\ref{ddg8}) vanishes, 
which means, in particular, that $\tilde e$ loses its
divergence-like $\xi$-symmetry that, as we explained, 
appears from the compensating mechanism between 
$\xi_{\bf{d-4}}^{a}$ and $\psi_{\bf{1}}^{a[d-3]}\,$ needed
for preserving the gauge condition $e_{\bf{d-3}}^{a}=0\,$. 
On the other hand, 
the last term on the right-hand side of (\ref{ddg4}) also disappears,
which entails that (\ref{ddg4}) contains not only 
constraint pieces, but also a first-order, 
Proca-like, dynamical equation.
As for the second special value $\alpha=3/2(-1)^d$,
the projections of $t_{\bf{2}}^{a[d-3]}$ that can be expressed in terms 
of $\omega_{\bf{1}}^{a[d-2]}$ from (\ref{ddg1}) 
cancel the $\omega$-term, 
when plugged into (\ref{ddg4}). Let us note, however, that 
both these cases still propagate the same number of degrees 
of freedom as dual graviton, which is ensured by the construction
and by inspection of the zero-form Weyl module that is left unchanged.
We do not consider these two special cases in more details here.

\section{Dualisation of arbitrary massless fields}
\label{sec:mixed}

Let us now discuss the dualisation of general 
massless mixed symmetry field described by the frame-like formulation.
As it was explained in Section \ref{umsf} the action for a massless 
spin-$\mathbf{Y}[h_1,h_2,h_3,\dots]$ field is given by 
\begin{equation}
\label{magain}
S=\langle de+\frac{1}{2}\sigma_-\omega|\omega\rangle \;,
\end{equation}
where the frame-like field $e$ is an $h_1$-form, valued in 
$\mathbf{Y}^1=\mathbf{Y}[h_2,h_3,\dots]$-shaped traceless tensors, 
while the spin-connection-like field $\omega$ is an $h_2$-form, valued in 
$\mathbf{Y}^2=\mathbf{Y}[h_1+1,h_3,\dots]$-shaped traceless tensors.
\vspace*{.3cm}

 As it can be noticed from the 
gravity dualisation examples, there are two different types of dualisations.
The first and the second dualisations of gravity are of the first and 
of the second types, respectively.
The first type of dualisation entails Hodge conjugation with the antisymmetric rank-$(d-2)$  tensor in the 
first column of the Young diagram representing the spin of 
the particle. 
This operation maps allowed Young diagram to the allowed ones on-shell.
Indeed, for the allowed Young diagram $h_1+h_2\leqslant d-2$ 
and $h_1\geqslant h_2\,$: 
after dualisation of the first column  one obtains a Young diagram with the first 
column of height $h'_1=d-2-h_1$, while the heights of other columns remain 
unchanged. 
It is easy to see that 
\begin{equation*}
h_1\geqslant h_2 \quad \Leftrightarrow \quad h_1'+h_2\leqslant d-2\;, 
\end{equation*}
\begin{equation*}
\quad h_1+h_2\leqslant d-2 \quad \Leftrightarrow \quad h_1'\geqslant h_2\;.
\end{equation*}
So, the first column, after dualisation, remains the highest one, and 
moreover the dual Young diagram is also allowed. 
\vspace*{.3cm}

In order to perform the first column off-shell dualisation, 
we add a torsion-like field $t$  being 
an $h_1+1$ differential form, valued in 
$\mathbf{Y}^1$-shaped traceless tensors. 
It is chosen such that 
its gauge symmetries of different levels can be used to gauge away the 
original frame-like field $e$ together with all its gauge symmetries. 
We also add the dual frame-like field $\tilde e$, valued in the same 
space $\mathbf{Y}^1$ as $t$ but carrying a differential form 
degree  $d-h_1-2\,$. 
The parent action is given by
\begin{equation}
\label{parentmix1}
S=\langle de+\frac{1}{2}\sigma_-\omega+t|\omega\rangle
+\int t\cdot d\tilde{e}\;,
\end{equation}
where ``$\,\cdot\,$'' implies that all the fiber indices are contracted
and of course, as always in this work, only the wedge product is 
used for multiplication of differential forms. 
The field $\tilde{e}$ can be treated as a Lagrange multiplier for the 
constraint $dt=0\,$. 
It can be solved in the form $t=d\beta$, 
implying that $t$ can be set to zero by gauge fixing, 
which shows that 
the action (\ref{parentmix1}) is equivalent to (\ref{magain}).
\vspace*{.3cm}

In order to construct the dual action one should 
gauge away the original frame-like $e$ 
field using the gauge parameters of $t$. 
Then $t$ can be completely expressed in terms 
of $\omega$. 
Plugging this back into the action we end up with an action 
formulated in terms of the dual frame-like field $\tilde{e}$ 
and the connection-like field $\omega\,$. 
To recast this dual action into the usual first-order form 
(\ref{magain})
one should define the dual connection-like field $\tilde\omega$  
as the Hodge dual in the first column, 
taken with respect to the fiber epsilon symbol with $d$ indices, 
of the original connection $\omega\,$.  
So, $\tilde\omega$ is an $h_2$-form transforming in the traceless 
$\mathbf{Y}[d-h_1-1,h_3,\dots]$-representation of the Lorentz algebra. 
The form degrees and fiber space types of $\tilde{e}$ and 
$\tilde{\omega}$ are exactly those (\ref{unfspace}) 
required in order to describe a spin-$\mathbf{Y}[d-h_1,h_2,h_3,\dots]$ 
particle in the first-order formulation \cite{Skvortsov:2008sh}. 
The parent action 
(\ref{parentmix1}) then reduces to (\ref{action}) for a
spin-$\mathbf{Y}[d-h_1,h_2,h_3,\dots]$ particle.
\vspace*{.3cm}

The second type of dualisation is a dualisation
which on-shell Hodge dualises any column 
-- including an empty column, which can formally be added 
to the right of the Young diagram -- of the Young diagram except
for the first one.
Suppose one is going to Hodge dualise the $i$th column of 
the traceless tensor of allowed form 
$\mathbf{Y}[h_1, h_2, \dots, h_i, \dots]\,$, 
so that one has $h_1+h_i\leqslant d-2\,$. 
After dulisation the $i$th column gives rise to the column with $h'_i=d-2-h_i$ boxes of the dual Young diagram.
Then
\begin{equation*}
h_1+h_i\leqslant d-2 \quad \Leftrightarrow \quad h_1\leqslant h'_i\;. 
\end{equation*}
So the column with height $h'_i$ appears to be the highest one in 
the dual Young diagram and therefore should 
be reshuffled to the first place on the left, 
such that the heights of the dual Young digram 
columns do not increase. 
Moreover
\begin{equation*}
h_1\geqslant h_i \quad \Leftrightarrow \quad h_1+ h'_i\geqslant d-2\;, 
\end{equation*}
which implies that the dual Young diagram is not allowed. 
It  means in turn that the dual tensor of ${so}(d-2)$
is not traceless, so to
describe the dual theories of this kind 
one needs an action of a form different from (\ref{action}).
\vspace*{.3cm}

In order to perform such a dualisation at the level of the action
we will generalise in a straightforward way the procedure followed 
in the previous section for the double-dual spin-2 field.
We first introduce an auxiliary field $t$,
being an $h_i+1$ form valued in traceless ${so}(d-1,1)$ 
tensors of shape 
$\mathbf{Y}^t=\mathbf{Y}[h_1,\dots,h_{i-1},h_{i+1}\dots]\,$, 
which can be obtained from 
the Young diagram characterizing the spin of the particle by 
cutting off the $i$th column. 
We also introduce  a dual frame-like field $\tilde{e}$
such that $t \cdot d\tilde{e}$ is a $d$-form scalar. 
In other words, $\tilde{e}$ it is a $d-h_i-2$ form valued in $\mathbf{{Y}}^t\,$.
The parent action is
\begin{equation}
\label{parentmix2}
S=\langle de+\frac{1}{2}\sigma_-\omega+\Sigma_- t|\omega\rangle+
\int (t\cdot d\tilde{e}+\alpha \,t^2)\;,
\end{equation}
where $\Sigma_{-}$ is an operator constructed from the product 
of $h_1-h_i$ background vielbeins and 
mapping $\mathbf{Y}^t$-shaped traceless tensors to 
$\mathbf{Y}^1$-shaped traceless tensors in fiber, which defines it  
uniquely,
while $\alpha \,t^2$ represents all the possible contractions,
each coming with an arbitrary coefficient, of 
two $t$ fields by means of $d-2h_i-2$ background 
vielbeins and the ${so}(d-1,1)$ epsilon tensor.  
Such contractions always exist for $h_1\geqslant h_i+1\,$. 
\vspace*{.3cm}

Performing the same manipulations as before one can show that  (\ref{parentmix2})
 is equivalent to (\ref{magain}).
To pass to the dual formulation we should gauge away the dynamical 
field of the original theory  (\ref{magain}). In the examples considered in Section \ref{fldg} the 
gauge parameter $\psi$ associated with the field $t$
 was used to completely gauge   away the 
frame-like field $e$ of the original theory. 
However, it is not possible in the general case.
Indeed, taking into account both base and fiber space indices, $e$ and 
$\psi$  carry the 
following tensor representations of the Lorentz group
\begin{equation*}
e: \qquad \mathbf{Y}[h_1]\otimes \mathbf{Y}[h_2,h_3\dots]\;,
\end{equation*}
\begin{equation*}
\psi: \qquad \mathbf{Y}[h_i]\otimes \mathbf{Y}[h_1,\dots,h_{i-1},h_{i+1},\dots]\;.
\end{equation*}
In general, these representations are different, so $\psi$ cannot be used to gauge away $e$ completely. 
Indeed, on the one hand we know from the frame-like formulation 
-- to which we borrow all the gauge symmetries for $e$ -- , that the action can be shown 
to contain only the component of $e$ transforming as a double-traceless
$\mathbf{Y}$ representation, where the double trace is taken with respect 
to four indices sitting in the same row. This is $\varphi^{_{\mathbf{Y}}}\,$, 
the Labastida metric-like gauge field off-shell. There are enough 
algebraic gauge symmetries for the frame field $e$ to reach that gauge.
\vspace*{.3cm}

On the other hand, it is clear that the parameter $\psi$  
contains and irreducible ${gl}(d)$ representation 
with shape $\mathbf{Y}\,$, 
but in general with more trace constraints than the one characterising
the Labastida field $\varphi^{_{\mathbf{Y}}}\,$, 
meaning that $\psi$ possesses less components than the Labastida 
field and hence cannot be used to completely gauge away 
the $e$ field inside the action.
This can be done, however, by resorting to the remaining 
differential gauge invariance of the Labastida field $\varphi^{_{\mathbf{Y}}}\,$ 
and reaching the gauge where $\varphi^{_{\mathbf{Y}}}\,$ becomes traceless, 
$\widehat{\varphi}^{{\,}_{\mathbf{Y}}}\,$, 
at the expense of leaving an action invariant under 
transverse gauge parameters, see \cite{Skvortsov:2007kz}
for the totally symmetric spin-$s$ cases. 
Then, at that stage, the gauge parameter $\psi$ can be used 
in order to completely gauge away the resulting traceless field $\widehat{\varphi}^{_{{\,}\mathbf{Y}}}\,$. 
\vspace*{.3cm}

The fact that the elimination of the original field from the parent action
requires the use of differential gauge symmetries may seem not elegant. 
To overcome this difficulty, one may introduce a set of auxiliary fields $t$
(instead of a single $t$-field), 
such that the gauge symmetries associated with them can be used
to eliminate the original frame field just by algebraic gauge shifts.
We leave this issue for further investigations.

\section{Conclusions}
\label{sec:concl}

In this paper, we performed a off-shell Hodge dualisation for massless 
mixed-symmetry fields in the Minkowski space of arbitrary dimension $d\,$. 
The dual fields are related on-shell by $so(d-2)$ 
Hodge conjugation on a group of indices associated with one column of 
the Young diagram describing the generalised spin of the initial field. 
We built the dual actions by introducing a parent action which, 
depending on the way one fixes gauges and eliminates fields by equations of motion, 
reduces to either the initial standard action \cite{Skvortsov:2008sh} 
or to the new, dual theory. The parent action procedure guarantees 
that both theories propagate the same number of degrees of freedom.
\vspace*{.3cm}

The frame-like approach has the advantage that it allows to promote 
the field equations to their unfolded formulation, and the latter formulation
requires the introduction of auxiliary fields that are precisely those needed
in order to build a frame-like parent action. 
The parent actions built within the frame-like approach are 
considerably simpler compared to their metric-like counterpart.
The frame-like action also makes the gauge symmetries manifest. 
\vspace*{.3cm}

As far as the counting of physical degrees of freedom is concerned, 
another great advantage of the unfolded formulation is brought by
the Weyl module representation which appears in the unfolded equations.
This representation contains an infinite set of zero-forms
that precisely carry the propagating degrees of freedom and therefore
makes their counting straightforward,  avoiding all the gauge-fixing
difficulties.
\vspace*{.3cm}

We start from the standard first order frame-like action \cite{Skvortsov:2008sh},
which on-shell describes irreducible tensors of $so(d-2)$ 
characterised by some Young diagram $\mathbf{Y}\,$. 
Performing the first-column off-shell dualisation of such a theory
in the way we proposed produces a dual theory which, on-shell
also gives an $so(d-2)$ traceless tensor characterised by a Young diagram
$\widetilde{\mathbf{Y}}\,$ related to $\mathbf{Y}$ by 
$so(d-2)$ Hodge dualisation in the first column. 
The dual action thereby obtained is the standard frame-like action \cite{Skvortsov:2008sh} 
for the dual field. 
\vspace*{.3cm}

On the other hand, dualising the inital action 
along the lines that we proposed on a column of $\mathbf{Y}\,$ 
which is \emph{not} the first one, 
we obtained a dual theory which describes, on-shell, 
a dual ${gl}(d-2)$-irreducible field which turns out to be
proportional to the metric tensor of $so(d-2)\,$.
We call such an on-shell field ``pure-trace''. 
The corresponding ${gl}(d)$-covariant field equations
can be expressed in terms of higher traces of the generalised
curvature tensor, and not via a single trace 
as is the case \cite{Bekaert:2003az,Bekaert:2006ix} 
for an on-shell gauge field which is not proportional to the
$so(d-2)\,$ metric tensor in the light-cone gauge.  
\vspace*{.3cm}

Such "higher-trace" theories had been studied on-shell in \cite{Hull:2000zn,Hull:2001iu,Bekaert:2002dt}
but so far, no ${gl}(d)$-invariant off-shell formulations had been found.
The present work together with \cite{Boulanger:2012df} fill this gap.
Indeed, the action \cite{Skvortsov:2008sh} is not suitable for such theories 
as, for the corresponding field of the dual types we considered, 
it does not propagate any pure-trace fields on-shell. 
\vspace*{.3cm}

The frame-like, dual actions producing a pure-trace field on-shell
contain two extra fields $(e,t)$ on top of the fields $(\tilde{e},\tilde{\omega})$
that one could expect to arise in a first-order approach.
We constructed and analysed in details the frame-like action for the double-dual graviton, 
as this case contains already all the features of the general, mixed-symmetry case. 
The double-dual graviton in $d$-dimensional Minkowski spacetime
is given on-shell by a $\mathbf{Y}[d-3,d-3]$-shaped 
tensor being $(d-4)$-fold pure-trace. In particular, it is shown 
that such theories admit exotic differential symmetries containing divergences of gauge parameters.
\vspace*{.3cm}

In the dualisation procedure considered in this paper, we replaced the 
lowest-grade frame-like field $e$ by the dual frame-like field $\tilde{e}\,$, 
but the first connection $\omega$ together with all the other higher-grade 
fields of the unfolded approach \cite{Skvortsov:2008vs} remain the same.
It would be  interesting to develop dualisations schemes which would involve 
non-trivial dualisations of some higher-grade fields.
\vspace*{.3cm}

One can consider the results obtained in this paper 
as a continuation of the program consisting in building
covariant actions for all the possible irreducible particles 
propagating freely in flat spacetime, in all the possible 
dual representations.
From this point of view 
it would be interesting to generalize the results of the paper to the 
multiple dualisation case. In general, one can study representations given 
by the arbitrary irreducible 
trace constraints, not necessarily obtained by the Hodge conjugation of 
$so(d-2)$-traceless on-shell tensors. 
One could also make the theory non-linear, along the lines 
of \cite{Nicolai:2000sc,Nicolai:2001sv,Nicolai:2003bp,deWit:2003ja,deWit:2008ta} 
in the spin-1 case, and \cite{Boulanger:2008nd} for the dual spin-2 case.

\section*{Acknowledgements}
\label{sec:Acknowledgements}

We are grateful to E. D. Skvortsov and P. Sundell for their comments. 
N.B. thanks the Erwin Schr\"odinger Institute in Vienna for kind hospitality.  
The work was supported in parts by an ARC contract No. AUWB-2010-10/15-UMONS-1.

\section{Appendix: Notation}

We deal with the $d$-dimensional 
Minkowski space parametrized by coordinates $x^{\mu}$.
The differential form indices are denoted by Greek letters $\mu,\nu, \dots$.
In each point of the space-time the vielbein $h_{\mu}^a(x)$ 
defines a local free falling basis with a flat metric 
$\eta^{ab}=\text{diag}(1,-1,\dots,-1)$, which is invariant tensor 
of $so(d-1,1)$. The 
tensor indices in this basis are denoted by Latin letters
from the beginning of the alphabet $a,b, \dots$  and 
often referred to as  fiber indices. Choosing Cartesian coordinates 
$h^a_{\mu}=\delta_{\mu}^a$ one identifies base and fiber indices. We also 
use Latin letters form the middle of the alphabet $m,n, \dots$ to denote tensor 
indices of the Wigner little group.

Differential form degree is often indicated as a lower index written in bold. 
The square bracket with the indices placed inside implies antisymmetrization 
of respective indices, 
while the round bracket denotes symmetrization. Both operations are supplied 
with overall factors making them projectors, e. g. 
$A^{[a}B^{b]}=1/2(A^aB^b-A^bB^a)$. For a group of $n$ (anti)symmetric indices 
we often use notation ($a[n]$) $a(n)\,$. 
We will also use the convention whereby tensors whose Lorentz indices 
are denoted by the same Latin letter are implicitly (anti)symmetrized 
on these indices. For example, $A^aB^a = A^{[a}B^{a]}\,$.  
It will be clear from the context whether one is working with the 
manifestly symmetric or antisymmetric convention. 
To indicate that tensor possesses 
a symmetry of a Young diagram 
written in a (anti)symmetric basis
we separate groups of (anti)symmetric indices
 by commas, e. g. $\mathbf{Y}[2,2,1]$-shaped 
tensor can be written either as $T^{a[2],b[2],c}$ in antisymmetric basis 
or as $T^{a(3),b(2)}$ in the symmetric basis.
If tensors indices are divided into groups and tensor 
does not possess any symmetries 
with respect to permutations of indices between the groups
these groups of indices are separated 
by vertical lines. For example, the differential form index of  
$V_{\bf{1}}^a$ can be transformed 
to the fiber one, which yields the second rank tensor $V^{m|a}$ with 
indefinite symmetry with  respect to permutation of indices.

We will also use notations
\begin{equation*}
h^{a[k]}=\underbrace{h^a\dots h^a}_{k} \quad 
\text{and} \quad H_{a[k]}=\epsilon_{a[k]b[d-k]}h^{b[d-k]},
\end{equation*}
where $\epsilon$ is totally antisymmetric rank $d$ tensor.

We also denote
\begin{equation*}
\delta_{n[d-k]}^{m[d-k]}=
\underbrace{\delta^{[m}_{n}\dots\delta^{m]}_{n}}_{d-k}, 
\quad
\sigma=\text{det}(\eta),
\end{equation*}
then
 \begin{equation}
 \label{epseps}
 \varepsilon^{a[k]m[d-k]}\varepsilon_{a[k]n[d-k]}
 =\sigma k!(d-k)!\delta_{n[d-k]}^{m[d-k]}.
 \end{equation}

\providecommand{\href}[2]{#2}\begingroup\raggedright\endgroup


\end{document}